# Study of 50 GeV proton beam focusing by novel crystal device

A.G.Afonin, V.I.Baranov, V.T.Baranov, G.I.Britvich, A.P.Bugorskiy,
M.K.Bulgakov, Yu.A.Chesnokov, P.N.Chirkov, A.A.Durum,
I.S.Lobanov, A.N.Lun'kov, A.V.Lutchev, V.A.Maisheev,
Yu.E.Sandomirskiy, A.V.Skleznev, A.A.Yanovich, I.A.Yazynin

*SRC IHEP, Protvino, Moscow region, 142281, Russia*

abstract

The bent crystals are applied on large accelerators to deflect particle beams in process of extraction and collimation. Recently the proposals of fixed target researches in the LHC are formulated. For realization of this program not only deflection but also focusing the LHC beam by bent crystals can be used. In the given work experimental results on 50 GeV proton beam focusing with the help of novel crystal device are reported. The positive property of this device is opportunity to work near the circulating beam of an accelerator, including the LHC.

At present, the collimation and extraction of a circulating beam using coherent phenomena in oriented crystals is examined at several large accelerators. Pioneering works [1–4] at the U-70 IHEP accelerator indicate that, in short bent silicon crystals, channeling can increase the efficiency of the extraction and collimation of the beam up to 85%. This possibility has been confirmed at the SPS (CERN) [5] and at the Tevatron (Fermilab) [6].

Recently in [7,8] the proposals of fixed target researches in the LHC are formulated. In these works it is marked, that the necessary extracted particle beams in TeV energy region can be received with the help of channeling in bent crystals. And, for performance of the program can be demanded not only deflection of particles, but also focusing them. In the given work experimental results on 50 GeV proton beam focusing with the help of novel crystal device are reported. The positive property of this device is opportunity to work near the circulating beam of an accelerator, including the LHC.

The principal opportunity of beam focusing by a bent crystal has been shown in [9] where the focusing effect was reached due to a difference of bending angles of particles in uniformly bent crystal with the oblique back end face. However used then device was inconvenient for applying near to a circulating beam of an accelerator because of the bend of a crystal was carried out by massive cylindrical mirrors. The new crystal device for beam focusing, which is convenient for installation in the accelerator, is shown on fig. 1.

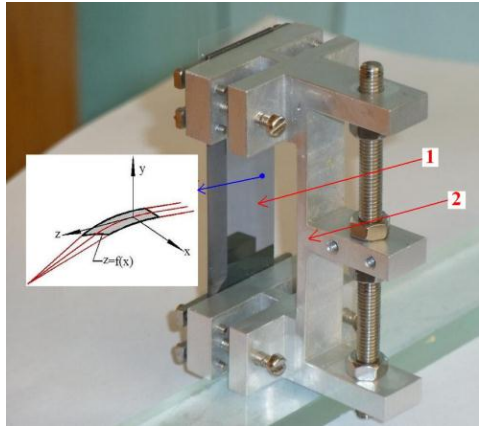

Fig. 1: The new crystal device for beam focusing . 1 - the bent crystal plate, 2 - the metal holder for a bend of a plate.

Extended on height the silicon plate (1) which have been cut out along crystallographic plane (111), is bent in a longitudinal direction with the help of the metal holder (2). The sizes of a plate are equal $x \times y \times z = 1 \times 70 \times 20$ mm$^3$. Due to anisotropic properties of material of a plate there is a transversal bend which is used for bending and focusing of a beam [10]. Trajectories of particles at movement through a crystal in channeling mode (in a horizontal plane) are shown on an insert in the left part of figure.

The form of a horizontal bend of a crystal (along coordinate z) in a focusing segment was measured by the laser device under the scheme described in ([11], page 86). The behavior of bending radius along coordinate $z$ can be approximated by the formula:

$$R(z) = \frac{R_o}{1 - Cz/L}$$

In this approximation the bending radius of a crystal increase with increasing of coordinate from 6 m in the beginning of a cut up to 30 m at the end of a cut. For focusing a beam in a point (without aberrations) it is necessary, that the bending angles of particles has been linearly connected to transversal coordinate $\varphi = \kappa x$. In case of a non-uniform bend of a crystal with arbitrary curvature the form of the oblique end face $z = f(x)$ is necessary to adjust. As shown in [10], focusing without aberrations occurs in a case

$$z = f(x) = \frac{L}{C}\left(1 - \sqrt{1 - \frac{2CR_o \kappa x}{L}}\right)$$

Where: $\kappa = (1 - C/2)L/(R_o d)$ at: $C = 0.8$, $L = 4.8$mm, $d = 0.8$mm, $R_o = 6$m.
Such form of a surface has been received at processing a silicon plate by the diamond blade, which is established on the programmable machine tool.

Experiment on focusing was carried out with 50 GeV proton beam of U-70 IHEP accelerator. The beam was extracted in Istra-Kristall setup which is located in 4A beamline. Apparatus opportunities of this setup are described in detail in [12]. The particle beam with the size $\sigma_x = 2$ mm and small angular divergence $\sigma_\alpha = 0.1$ mrad was directed on a crystal established in goniometer. Intensity of the incident beam and the beam deflected by a crystal on the angle of (2.3-2.8) mrad was measured by scintillator counters. At optimum orientation the crystal deflected ~ 3 % of particles of a direct beam that corresponds to calculations as the critical channeling angle is equal 0.03 mrad for particles with 50 GeV energy in planar channel Si (111). The effect of focusing has been registered with the help of nuclear photoemulsions. Few layers of emulsions were settled down on various distance from a crystal. The Fig. 2 illustrates effect of focusing of a beam on distance of $l_F = 1/\kappa = 1.7$ m from a crystal.

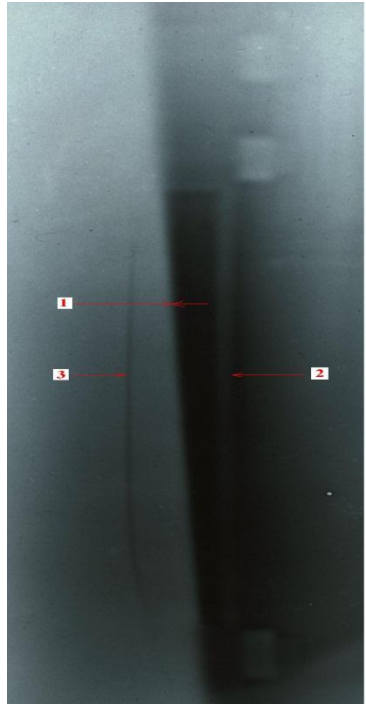

Fig. 2: the Image of a particle beam on emulsion layer located near to focus. 1-border of collimator, 2-the shadow of a crystal, 3-the focused beam.

With the help of extended lead collimator the sharp border of an incident particle beam is generated which is well visible in the center of photo. The focusing crystal is placed in an intensive part of a beam, and its shadow is reflected on the right in rays of an incident beam as a strip with precise borders. The beam deflected and focused by a crystal is seen at the left in a shadow of collimator as a narrow vertical line. Thus, it is evidently presented, as the beam equal on the size to thickness of a crystal (1 mm) is deviate and compressed in a line in width 0.22mm (FWHM). With the help of digital scanning the image (the strip few millimeters on a vertical is allocated) the horizontal profile of the whole beam (fig. 3a), and also a profile of the focused beam (fig. 3b) is received.

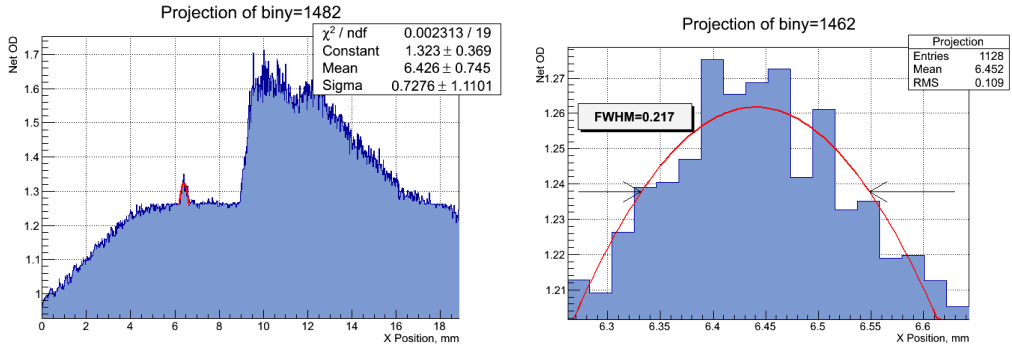

Fig. 3: **a** - a horizontal profile of the whole beam, there are visible: border of collimator in the center, shadow from a crystal on the right, and the focused beam at the left;
**b** - a horizontal profile of the focused beam.

Processing of the image was carried out with the help of the digital high resolution scanner ( 10 microns). On fig. 4 it is shown the measured beam envelope, constructed by results of processing of few emulsion layers on the digital scanner. The expected theoretical curve, with the account and without taking into account substance on a way of particles from a crystal before focus 1.7m (total amount of substance is about 1 $g/sm^2$) is shown also.

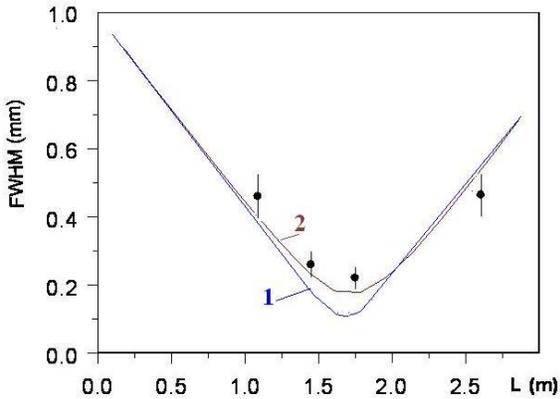

Fig. 4: Beam envelope (the size of a beam depending on distance from a crystal). Points - experiment. A curve 1 - calculation for transportation of particles in vacuum, a curve 2-calculation for real conditions in view of scattering on air and emulsion layers.

Apparently from figure, the good consent of parameters of focusing with expected value is observed. The size of a beam in focus within the limits of 10 % coincides with results of the modeling, taking into account real conditions of carrying out of experiment.

The focusing property of the developed device can be applied on the LHC or the other accelerator of high energy to research of low-angular processes. The crystal can be align on a fix target by focusing end face, as shown in fig. 5. Rotating the crystal around of an axis O, one can deflect the particles from the target aside from adverse background area near the circulating beam. I.e. the role of a crystal consists in creation of clean conditions for registration of the necessary particles. Other motive of application of such scheme is reception of a secondary particle beam of high energy by rather simple way.

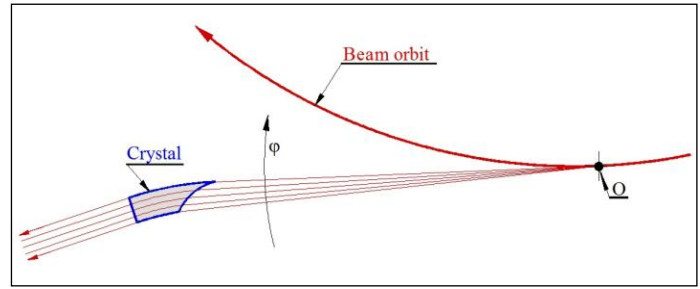

Fig. 5: An example of application of a focusing crystal for research of low-angular processes. The same scheme can be a source of a parallel beam of secondary particles for the wide physical researches suggested in [7,8].

Work is supported by IHEP directorate, and also the Russian fund of basic researches (the project 12-02-91532).